\documentclass[12pt,letterpaper]{article}
\usepackage[pdftex]{graphicx,color}
\usepackage{hyperref}
\input{epsf}
\usepackage{amsmath,amssymb}
\usepackage[dvips,letterpaper,text={6.5in,9in}]{geometry}
\usepackage{fancyhdr}
\usepackage{verbatim}

\newcommand\ltap{\
  \raise.3ex\hbox{$<$\kern-.75em\lower1ex\hbox{$\sim$}}\ }
\newcommand\gtap{\
  \raise.3ex\hbox{$>$\kern-.75em\lower1ex\hbox{$\sim$}}\ }

\newcommand\simge{\mathrel{%
   \rlap{\raise 0.511ex \hbox{$>$}}{\lower 0.511ex \hbox{$\sim$}}}}
\newcommand\simle{\mathrel{
   \rlap{\raise 0.511ex \hbox{$<$}}{\lower 0.511ex \hbox{$\sim$}}}}

\newcommand{\slashchar}[1]%
        {\kern .25em\raise.18ex\hbox{$/$}\kern-.75em #1}
\def\lsim{\mathrel{\raise.3ex\hbox{$<$\kern-.75em\lower1ex\hbox{$\sim$}}}}
\def\gsim{\mathrel{\raise.3ex\hbox{$>$\kern-.75em\lower1ex\hbox{$\sim$}}}}

\newcommand\CB{{\cal B}}

\newcommand\CH{{\cal H}}

\newcommand\CM{{\cal M}}

\newcommand\CO{{\cal O}}

\newcommand\be{\begin{equation}}
\newcommand\ee{\end{equation}}
\newcommand\bea{\begin{eqnarray}}
\newcommand\eea{\end{eqnarray}}
\newcommand\ba{\begin{array}}
\newcommand\ea{\end{array}}
\newcommand\nn{\nonumber}

\newcommand\dagg{\dagger}

\newcommand\mev{{\rm MeV}}
\newcommand\gev{{\rm GeV}}
\newcommand\tev{{\rm TeV}}

\newcommand\ellm{\ell^-}

\newcommand\ellp{\ell^+}

\newcommand\mum{\mu^-}

\newcommand\mup{\mu^+}

\begin{document}

\title{
\vskip -15mm
\begin{flushright}
 \vskip -15mm
 {\small CERN-PH-TH-2015-153\\
   LAPTH-037/15\\
 }
 \vskip 5mm
 \end{flushright}
{\Large{\bf Charged-Lepton Mixing and Lepton Flavor Violation}}\\
} \author{
  {\large Diego Guadagnoli$^{1}$\thanks{diego.guadagnoli@lapth.cnrs.fr}}\, and
  Kenneth Lane$^{2,3}$\thanks{lane@physics.bu.edu}\\
{\large $^{1}$Laboratoire d'Annecy-le-Vieux de Physique Th\'eorique} \\
{\large UMR5108\,, Universit\'e de Savoie Mont Blanc et CNRS} \\
{\large B.P.~110, F-74941, Annecy-le-Vieux Cedex, France}\\
{\large $^{2}$Department of Physics, Boston University}\\
{\large 590 Commonwealth Avenue, Boston, Massachusetts 02215}\\
{\large $^{3}$CERN Theory Division}\\
{\large CH-1211, Geneva 23, Switzerland}
}
\maketitle

\begin{abstract}
  
  We present a model for calculating charged-lepton mixing matrices. These
  matrices are an essential ingredient for predicting lepton flavor-violating
  rates in the lepton number nonuniversal models recently proposed to explain
  anomalies in $B$-meson decays. The model is based on work on ``constrained
  flavor breaking'' by Appelquist, Bai and Piai relating the charged-lepton
  mass matrix, $\CM_\ell$, to those for the up and down-type quarks,
  $\CM_{u,d}$.  We use our recent model of lepton nonuniversality to
  illustrate the magnitudes of flavor-violating $B$-decay rates that might be
  expected. Decays with $\mu\tau$ final states generally have the highest
  rates by far.

\end{abstract}


\newpage

The LHCb Collaboration has reported several features of $B$-meson decays
involving $b \to s \ellp\ellm$ transitions that consistently point to a
departure from the Standard Model (SM) of particle physics:

\begin{itemize}

\item The ratio $R_K$ of the decay rates of $B^+ \to K^+ \ellp\ellm$ for
  $\ell = \mu,e$~\cite{Aaij:2014ora}
\be\label{eq:RK}
R_K \equiv \frac{\CB(B^+ \to K^+ \mu^+\mu^-)}{\CB(B^+ \to K^+ e^+e^-)} =
0.745^{+0.090}_{-0.074}\,{\rm (stat)}\pm 0.036\,{\rm (syst)}\,.
\ee
This result is a 2.6$\sigma$ deficit from the standard model (SM) prediction,
$R_K = 1+\CO(10^{-4})$~\cite{Bobeth:2007dw,Bouchard:2013mia,Hiller:2003js}.

\item The direct measurement~\cite{Aaij:2014pli},
\be\label{eq:NewBRKmumu}
\CB(B^+ \to K^+\mu^+\mu^-)_{[1,6]} = (1.19 \pm 0.03 \pm 0.06)\times 10^{-7}\,.
\ee
This is about $30\%$ lower than the SM prediction, $\CB(B^+ \to K^+ \mu^+
\mu^-)^{\rm SM}_{[1,6]} = (1.75^{+0.60}_{-0.29}) \times
10^{-7}$~\cite{Bobeth:2011gi,Bobeth:2011nj,Bobeth:2012vn}. 

\item The observable $P'_5$ in $B^0 \to K^{*0} \mu^+ \mu^-$ angular
  distributions exhibits a deficit in two bins, quantified by LHCb as
  2.9$\sigma$ for each bin~\cite{LHCb:2015iha}. However, the theoretical
  error is debated~\cite{Descotes-Genon:2013wba,Descotes-Genon:2014uoa,
    Jager:2014rwa}.

\end{itemize}  

\noindent These three measurements were made in the low $q^2= M^2_{\ell\ell}$ 
region of $1.0$--$6.0\,\gev^2$, away from charmonium resonances in the
$\ellp\ellm$ spectrum.

\begin{itemize}

\item  The joint CMS--LHCb measurement~\cite{CMS:2014xfa}
\be\label{eq:Bsmumu}
\CB(B_s \to \mu^+\mu^-)_{exp} = (2.8^{+0.7}_{-0.6})\times 10^{-9} =
(0.76^{+0.20}_{-0.18})\times \CB(B_s \to \mu^+\mu^-)_{SM}\,.
\ee
Although this is consistent with the SM prediction~\cite{Bobeth:2013uxa}, the
central value is about 25\% low --- as it is for
$R_K$~\cite{Glashow:2014iga}.

\end{itemize}

\noindent The $R_K$-measurement suggests lepton nonuniversality (LNU) occurs
in $b\to s\ellp\ellm$ transitions; the other measurements are consistent in
magnitude and sign. It is no wonder, then, that they have inspired a number
of LNU models of new physics (NP) above the electroweak energy scale,
involving the exchange of multi-TeV
particles~\cite{Altmannshofer:2013foa,Gauld:2013qba, Buras:2013qja,
  Gauld:2013qja,Buras:2013dea,Altmannshofer:2014cfa, Hiller:2014yaa,
  Ghosh:2014awa,Altmannshofer:2014rta, Hiller:2014ula,
  Gripaios:2014tna,Bhattacharya:2014wla,Glashow:2014iga,
  Crivellin:2015mga,Crivellin:2015lwa,Niehoff:2015bfa,Sierra:2015fma,
  Celis:2015ara, Becirevic:2015asa,Varzielas:2015iva,Boucenna:2015raa,
  Crivellin:2015era,Lee:2015qra,Alonso:2015sja,Greljo:2015mma,
  Calibbi:2015kma}.

LNU interactions at high energy are accompanied by LFV interactions unless the 
leptons involved are chosen to be mass eigenstates~\cite{Glashow:2014iga}. Such 
a choice is an act of fine tuning in the absence of a dynamical or symmetry 
mechanism justifying it.\footnote{Attempts in this direction are in 
Refs.~\cite{Celis:2015ara,Alonso:2015sja}.}
Further, since charged leptons (and quarks) are massless at
$\Lambda_{LNU}$, far above the weak scale, it is difficult to understand
the motivation or need for flavor-invariant Yukawa couplings there.
If the anomalies reported by
LHCb hold up, LFV decays such as $B \to K^{(*)} \mu e$ and $B \to K^{(*)} \mu
\tau$ should occur at rates much larger than in the SM due to tiny neutrino
masses alone. The purpose of this paper is to present a model for estimating
these and other LFV rates implied by new LNU interactions.

LHCb data suggest that LNU is affecting muons but not electrons. To describe
this, a simple model was adopted in Ref.~\cite{Glashow:2014iga} (hereafter
referred to as GGL) in which a heavy $Z'$ boson couples only to
third-generation quarks and leptons, namely,
\be\label{eq:HNP}
\CH_{NP} = G\, \bar b'_L\gamma^\lambda b'_L \, \bar\tau'_L \gamma_\lambda
\tau'_L\,.
\ee
This chiral structure is consistent with $B$-decay data which is well fit if
the SM {\rm and} NP contributions to the $b \to s \ellp\ellm$ interaction are
a product of left-handed currents (LL)~\cite{Hiller:2014yaa,Ghosh:2014awa,
  Altmannshofer:2014rta,Hurth:2014vma}. In Eq.~(\ref{eq:HNP}), $G =
g^2_{Z'}/M^2_{Z'} = 1/\Lambda_{NP}^2 \ll G_F$ is a new Fermi constant. The
primed fields refer to the gauge basis, the one in which the charged weak
currents are generation-universal.\footnote{This interaction has been
  extended in Ref.~\cite{Bhattacharya:2014wla} to the $SU(3)\otimes
  SU(2)\otimes U(1)$-invariant form it must have if $\Lambda_{NP} \gg
  \Lambda_{EW}$~\cite{Alonso:2014csa}, and used to provide a simultaneous
  explanation for $R_K$ and $R(D^{(*)})$. A consistent gauge model must also
  be anomaly-free. These extensions are not needed in this paper.} They are
related to mass-eigenstate (unprimed) fields by unitary matrices $U_L^d$ and
$U_L^\ell$:
\be\label{eq:Umatrices}
b'_L \equiv d'_{L3} = \sum_{i=1}^3 U^d_{L3i} \, d_{Li}\,,\quad
\tau'_L \equiv \ell'_{L3} = \sum_{i=1}^3 U^\ell_{L3i}\, \ell_{Li}\,.
\ee
The interaction responsible for the discrepancies in $R_K$, $B^+ \to K^+
\mup\mum$, $B_s \to \mup\mum$ and $P'_5$ is then
\be\label{eq:bsmmNP}
\CH_{NP}(\bar b \to \bar s \mu^+\mu^-) = G\left[U^{d*}_{L33} U^d_{L32}\vert
  U^\ell_{L32}\vert^2 \, \bar b_L \gamma^\lambda s_L \, \bar \mu_L
  \gamma_\lambda \mu_L + {\rm H.c.}\right].
\ee

In GGL we said that the hierarchy of the Cabibbo-Kobayashi-Maskawa (CKM)
matrix $V_{CKM} = U_L^{u\dagg} U_L^d$ for quarks and the apparent preference
of the new physics for muons over electrons suggest that
$|U^{d,\ell}_{L31}|^2 \ll |U^{d,\ell}_{L32}|^2 \ll |U^{d,\ell}_{L33}|^2
\simeq 1$. Then, in order that this Hamiltonian deplete the SM contribution,
we assumed $GU^{d*}_{L33} \,U^d_{L32} < 0$. This sign choice is correct if
$U^d_L \simeq V_{CKM}$. In truth, however, we know little about $U^d_L$ and
$U^\ell_L$ other than $V_{CKM} = U_L^{u\dagg} U_L^d$ and that $U^\ell_L$
plays a similar part in the less well-measured PMNS matrix,
$V_{PMNS}$~\cite{Agashe:2014kda}.

Experimentalists need better targets. A recent paper by Boucenna, Valle and
Vicente~\cite{Boucenna:2015raa} made a first stab at this by assuming
$U_L^\ell = V_{PMNS}$. In our opinion, $V_{PMNS} = U_L^{\nu \dagg}U^\ell_L$
seems likely to be strongly influenced by the unknown neutrino mixing
matrix. Ref.~\cite{Varzielas:2015iva} discussed lepton flavor mixing in the
context of LNU due to a leptoquark interaction~\cite{Hiller:2014yaa}. While
we will present results for the $Z'$-induced $\CH_{NP}$, our scheme for
charged-lepton mixing is independent of the dynamical nature of LNU and LFV.

The model we propose for calculating charged-lepton mixing matrices,
$U^\ell_{L,R}$, is based on a recent paper on ``constrained flavor breaking''
by Appelquist, Bai and Piai (ABP)~\cite{Appelquist:2015mga}. They assumed a
global constrained flavor symmetry group, $SU(3)^3$, broken by just two
Yukawa spurions. This implies one equation among the three Yukawa matrices,
$Y_u, Y_d, Y_\ell$. They are related to the quark and charged-lepton mass
matrices $\CM_{a}$ in
\be\label{eq:Hmass}
\CH_{mass} = \sum_{i,j=1}^3 \left[\bar u'_{Li} \CM_{u\,ij} u'_{Rj} +
\bar d'_{Li} \CM_{d\,ij} d'_{Rj}+ \bar \ell'_{Li} \CM_{\ell\,ij}
\ell'_{Rj} + H.c.\right]
\ee
by $\CM_{a} = Y_a v/\sqrt{2}$ where $v$ is the Higgs vacuum expectation
value. The matrices $U^a_{L,R}$ bring these to real, diagonal, positive form:
\be\label{eq:Mdiag}
\widehat\CM_a \equiv \CM_{a,\,diag} = U^{a\dagg}_L \CM_{a} U^a_R, \qquad
(a=u,d,\ell)\,. 
\ee
 \begin{figure}[!t]
   \begin{center}
     \vspace*{-0.5in}
     \includegraphics[width=2.00in, height = 2.00in, angle=0]
     {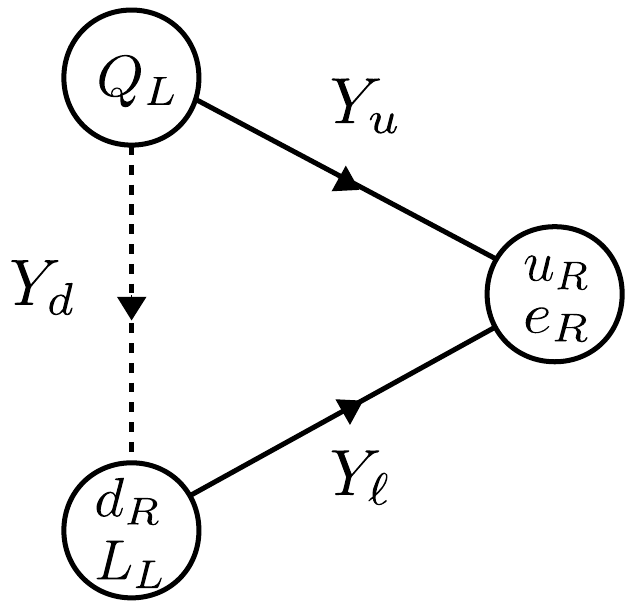}
     \caption{Moose diagram for the ABP model of constrained flavor breaking,
       Ref.~\cite{Appelquist:2015mga}. The solid links are the input Yukawa
       matrices chosen by ABP (our case~A), the dashed link is then
       predicted.}
     \label{fig:moose}
   \end{center}
 \end{figure}

 The ABP model is based on the moose diagram of Fig.~1. Requiring (a) that
 the quark doublet and singlet fields, $Q_L, u_R$ and $d_R$, must be assigned
 to different $SU(3)$'s (to have realistic masses and $V_{CKM}$); (b) that
 $L_L$ and $e_R$ likewise be assigned to different $SU(3)$'s; and (c) that
 $L_L$ and $e_R$ be assigned to $SU(3)$ groups other than $Q_L$'s (to avoid
 $\widehat \CM_\ell \propto \widehat \CM_u$ or $\widehat \CM_d$), leaves six
 possibilities~\cite{Appelquist:2015mga}. The one chosen by ABP is depicted
 Fig.~\ref{fig:moose}. Having taken $Y_u$ and $Y_\ell$ independent, $Y_d$ is
 predicted up to a constant, $\eta$:
\be\label{eq:YABP} Y_d = \eta Y_u Y_\ell^\dagg\,.
\ee
Equivalently, $\CM_{d} = \widehat \eta \CM_{u} \CM^\dagg_{\ell}$.

As ABP were interested only in masses of the charged leptons, not their
mixing, they did not consider assignments differing by an interchange of
$L_L$ and $e_R$ which just interchanges $Y_\ell$ and $Y^\dagg_\ell$. But,
this swaps $\CM_{\ell}$ and $\CM^\dagg_{\ell}$ and, hence, $U^\ell_L$ with
$U^\ell_R$, resulting in different mixing factors for LNU/LFV processes. ABP
rejected choosing $Y_u$ and $Y_d$ as the flavor-breaking spurions because that
implies $Y_\ell = \eta Y^\dagg_d Y_u$ and, hence, unrealistic charged-lepton
masses. We agree. They also rejected the possibility $Y_u = \eta Y_d Y_\ell$,
arguing that it has difficulty obtaining a large enough top Yukawa coupling.
But $\eta$ is a free parameter of unknown origin, and there is considerable
freedom in choosing the textures for the $Y_a$, so we will consider this
case. We will see it is closely related to the case ABP considered. Thus, we
consider four cases. Written as a relation from which $\CM_{\ell}$ is
determined, they are (ignoring the dimensionful $\eta$-factor):
\be\label{eq:Mlcases}
\CM_{\ell} = \left\{\begin{array}{lr}
   (\CM^{-1}_{u}\, \CM_{d})^\dagg   \qquad & {\rm (A)}\\
   \CM^{-1}_{u}\, \CM_{d}           \qquad & {\rm (B)}\\
   \CM^{-1}_{d}\, \CM_{u}           \qquad & {\rm (C)}\\
   (\CM^{-1}_{d}\, \CM_{u})^\dagg   \qquad & {\rm (D)}
\end{array}\right.
\ee

By a special choice of quark bases, neither the gauge nor mass
bases defined above, ABP obtained the mass matrix and, hence, mass ratios for
the charged leptons from Eq.~(\ref{eq:YABP}) in terms of the known quark
masses and CKM matrix elements~\cite{Agashe:2014kda}. No assumption of
particular quark mass textures was required. While their results are not in
agreement with data, they are not all that bad, so there is promise in their
approach.  But, the matrices diagonalizing their lepton mass matrix are not
transformations from the gauge to the mass basis, and thus cannot be used for
turning $\CH_{NP}$ in Eq.~(\ref{eq:HNP}) into predictions of LFV rates.  For
that, we need specific textures for $\CM_{u,d}$ in the gauge basis, ones that
provide a reasonable account of the quark masses {\em and} $V_{CKM}$.

Fortunately, quark mass textures good enough for our purpose exist; see,
e.g., Ref.~\cite{Feruglio:2015jfa}. We use ones developed in connection with
a scenario for solving the strong-CP problem in QCD~\cite{Martin:2004ec}. In
this scenario, the phases in $\CM_{u,d}$ are rational multiples of $\pi$ so
that they easily satisfy
\be\label{eq:thetabar}
\bar\theta \equiv \arg\det\CM_q = \arg\det\CM_u + \arg\det\CM_d = 0\,.
\ee
The mass matrices are $\CM_{q=u,d} = U_L^q \widehat\CM_q U_R^{q\dagg}$ where
we use $\overline{\rm MS}$ quark masses renormalized at the top-quark pole
  mass $M_t = 173.5\,\gev$, with eigenvalues $\widehat \CM_u={\rm
    diag}(0.00126,0.611,163.5)\,\gev$ and $\widehat \CM_d={\rm
    diag}(0.00264,0.0522,2.72)\,\gev$.\footnote{We also use $m_b(m_b) =
  4.18\,\gev$, $m_c(m_c) = 1.275\,\gev$, $m_s(2\,\gev)=95\,\mev$,
  $m_d(2\,\gev) = 4.8\,\mev$ and $m_u(2\,\gev) =
  2.3\,\mev$~\cite{Agashe:2014kda}.} Then:
\bea\label{eq:Qmasses}
\CM_{u} &=& \left(\begin{array}{ccc} (0,0) & (0.01038,-2\pi/3) & (0,0)\\
    (0.1325,0) & (0.5964,0) & (0,0)\\ (0,0) & (0,0) & (163.5,0)\\
\end{array}\right)\,, \\
\CM_{d} &=& \left(\begin{array}{ccc} (0,0) & (0.01112,0) & (0.01322,0) \\
    (0.01013,\pi/3) & (0.05012,0) & (0.1127,\pi/3)\\ (0,0) & (0,0) &
    (2.721,\pi/3)\\
\end{array}\right)\,.
\eea
The notation is $(|\CM_{q,\,ij}|,{\rm arg}(\CM_{q,\,ij}))$. The motivation
for these mass textures is explained in Appendix~B of
Ref.~\cite{Martin:2004ec}.  Note that, since quark masses are
multiplicatively and {\em universally} renormalized above $M_t$,
the lepton mass textures in Eq.~(\ref{eq:Mlcases}) are insensitive to QCD
running from $M_t$ to $\Lambda_{NP}$.  The CKM matrix obtained by
diagonalizing $\CM_{u,d}$, removing its unphysical phases and casting it in
standard form~\cite{Harari:1986xf}, is

\be\label{eq:VCKM}
V_{CKM} = U_L^{u\dagg} U_L^d = 
\left(\begin{array}{ccc} (0.976,0) & (0.216,0) & (0.0045,-0.978)\\
                         (0.216,\pi) & (0.976,0) & (0.0415,0)\\ 
                         (0.0075,-0.516) & (0.410,3.161) & (0.999,0)\\ 
\end{array}\right)\,.
\ee
This reproduces measured CKM matrix entries to within a few per cent, except
for $V_{ub}$ and $V_{td}$ which are within 20\%~{\cite{CKM}}.

The $U^\ell_{L,R}$ are obtained (up to a diagonal matrix of pure phases) by
diagonalizing
\be\label{eq:Msql}
\CM^2_{\ell,\,LL} = \CM_\ell \CM_\ell^\dagg \,\,\, {\rm and} \,\,\, 
\CM^2_{\ell,\,RR} = \CM_\ell^\dagg \CM_\ell \,.
\ee
For cases A and C (and for cases B and D)
\be\label{eq:ACrelation}
\CM^2_{\ell,\,LL}(C) = \CM^{-2}_{\ell,\,LL}(A) \,.
\ee
Therefore, the (dimensionless) eigenvalues of $\CM^2_{\ell,\,LL}(C)$ are the
inverses of those of $\CM^2_{\ell,\,LL}(A)$, i.e.,
\be\label{eq:MlACevs}
(m^2_\tau,m^2_\mu,m^2_e)_C = (m^{-2}_e, m^{-2}_\mu, m^{-2}_\tau)_A\,,
\ee
and $U^\ell_{L,R}(C)$ are the same as $U^\ell_{L,R}(A)$ {\em with their first
  and third columns interchanged.}

The mass-squared matrices for case~A are
\bea\label{eq:MsqA}
\CM^2_{\ell,\,LL}(A) &=&
\left(\begin{array}{ccc} 
    (0.1218,0) & (3.191,2.50) & (3.413,2.57)\\
    (3.191,-2.50) & (83.84,0) & (89.69,0.0716)\\ 
    (3.413,-2.57) & (89.69,-0.0716) & (95.98,0)\\ 
\end{array}\right)\,, \\
[0.2cm]
\CM^2_{\ell,\,RR}(A) &=&
\left(\begin{array}{ccc} 
    (171.3,0) & (38.40,-3.08) & (0.1589,-1.96)\\
    (38.40,3.08) & (8.617,0) & (0.03660,1.11) \\ 
    (0.1589,1.96) & (0.03660,-1.11) &  (2.77 \times 10^{-4},0)\\
\end{array}\right)\,.
\eea

The predicted and measured ratios of the lepton masses are
\bea\label{eq:mlABCD}
m_e/m_\tau &=& 1.53\times 10^{-4}\,,\,\,m_\mu/m_\tau = 0.00802
 \qquad  {\rm (cases \,\, A,B)}\,;\nn\\
m_e/m_\tau &=& 1.53\times 10^{-4}\,,\,\,m_\mu/m_\tau = 0.0191 \qquad {\rm
 \,\,\,(cases\,\, C,D)}\,;\\
m_e/m_\tau &=&  2.88\times 10^{-4}\,,\,\,m_\mu/m_\tau = 0.0595 \qquad \,\,\,{(\rm
  Ref.~\cite{Agashe:2014kda})}\,.\nn
\eea
%
The predicted ratios are not great, but they do exhibit a qualitatively
correct hierarchy. Different quark mass textures will lead to different
ratios.\footnote{We have reproduced the results of
  Ref.~\cite{Appelquist:2015mga} using their $\CM_\ell$. The magnitudes of
  the corresponding $U^\ell_{L,R}$ matrix elements are similar to those
  in Eqs.~(\ref{eq:UlL},\ref{eq:UlR}).}

The lepton mixing matrices for this case are
 \bea
 \label{eq:UlL}
 U^\ell_L(A) &=& \left(\begin{array}{ccc}
     (0.9808,1.325) & (0.1935,-1.945) & (0.02597,3.026) \\ 
     (0.1515,1.065) & (0.7149,0.8359) & (0.6826,0.5264) \\
     (0.1231,-2.372) & (0.6719,-2.369) & (0.7304,0.4548) \\ 
 \end{array}\right)\,, \\
 \label{eq:UlR}
 U^\ell_R(A) &=& \left(\begin{array}{ccc} 
     (0.02203,2.337) & (0.2176,-0.7759) & (0.9758,-0.4548) \\ 
     (0.1016,2.303) & (0.9705,-0.8359) & (0.2187,2.627) \\
     (0.9946,-1.325) & (0.1039,-1.321) & (0.9062\times 10^{-3},1.502) \\
 \end{array}\right)\,,
 \eea
%
 where phases have been chosen to make $\widehat \CM_{\ell}$ real and
 positive (see Eq.~(\ref{eq:Msql})). The columns of $U^\ell_L(A)$ are the
 orthonormal eigenvectors $v_{e_L}, v_{\mu_L}, v_{\tau_L}$ of
 $\CM^2_{\ell,LL}(A)$ with rows labeled by $e', \mu', \tau'$, and similarly
 for $U^\ell_R(A)$. For the hermitian conjugate case, with $\CM_\ell(B) =
 \CM_u^{-1} \CM_d$, the mixing matrix $U^\ell_L(B) = U^\ell_R(A)$. The number
 of physical phases in $U^\ell_L$ depends on the nature of the neutrino
 sector, whether Dirac or Majorana. These phases may induce new sources of CP
 violation in decay, but only by interfering with SM amplitudes. Since LFV
 processes have at most tiny SM amplitudes, their rates involve only absolute
 values of $U^\ell_{L,R}$ elements.
 
 These mixing matrices, or ones developed from other quark-mass textures and
 the ABP ansatz, can be used to predict LNU and LFV rates in any NP model of
 these processes. For our $\CH_{NP}$, Eq.~(\ref{eq:HNP}), the elements of
 interest in $U^\ell_L$ are the third row, $v_{\tau'_L}$. In particular, the
 amplitudes for $B \to K^{(*)}\ell^+_i \ell^-_j$ and $B_s \to \ell^+_i
 \ell^-_j$ involve $U^{\ell *}_{3i} U^\ell_{3j}$. In case~A, $|U^\ell_{L,32}|
 \simeq |U^\ell_{L,33}|\simeq 1/\sqrt{2} \gg |U^\ell_{L,31}|$, contrary to
 our naive expectation that these matrices have a CKM-like
 hierarchy~\cite{Glashow:2014iga}. Even more surprising $|U^\ell_{R,31}| \gg
 |U^\ell_{R,32}| \gg |U^\ell_{R,33}|$. Note that this means that the
 $U^\ell_{L,3i}$ of case~D are CKM-like, as naively expected.  These features
 are a consequence of the block-diagonal $\CM^2_{\ell,LL}$ and
 $\CM^2_{\ell,RR}$, the latter exhibiting an extreme example of
 level-crossing. In turn, these trace back to the textures of $\CM_u$ and
 $\CM_d$ (and the ABP ansatz). $\CM_u$ is $(2\times 2) \oplus (1\times 1)$
 block-diagonal and employs a see-saw to make $m_u \ll m_c$ without an
 $\CO(m_u)$ matrix element. Approximately the same structure in $\CM_d$ plus
 $U^u_L \simeq 1$ lead to the famous relation $\tan\theta_C \cong \theta_{12}
 \simeq \sqrt{m_d/m_s}$ in $V_{CKM}$.

 Finally, we apply these results to our model ``third-generation''
 Hamiltonian, Eq.~(\ref{eq:HNP}), and evaluate branching ratios for the LFV
 processes $B \to K^{(*)} \ell^\pm_i \ell^\mp_j$ and $B_s \to \ell^+_i
 \ell^-_j$. Since these rates will be proportional to $|U^{\ell *}_{L3i}
 U^\ell_{L3j}|^2$, both lepton charge assignments may be combined. The first
 order of business is to note that cases~B and~C are excluded in our
 model. In those cases, $|U^\ell_{L31}|^2$ is not much smaller than
 $|U^\ell_{L32}|^2$, implying $R_K \simge 1$.

 For $\ell_i \ne \ell_j$, our model implies (summing over both lepton charge
 modes)
\bea
\label{eq:LFVBrates}
\frac{\CB(B \to K^{(*)} \ell^\pm_i \ell^\mp_j)}
{\CB(B \to K^{(*)} \mu^+\mu^-)} &\simeq&
2\rho_{NP}^2\left|\frac{U^{\ell *}_{L3i}
    U^\ell_{L3j}}{|U^\ell_{L32}|^2} \right|^2\,,\\
\label{eq:LFVBsrates}
\frac{\CB(B_s \to \ell^\pm_i \ell^\mp_j)}{\CB(B_s \to \mu^+\mu^-)} &\simeq&
2\rho_{NP}^2\left|\frac{U^{\ell *}_{L3i}
    U^\ell_{L3j}}{|U^\ell_{L32}|^2} \right|^2 \left[\frac{m_i^2 + m_j^2 -
    (m_i^2-m_j^2)^2/M_{B_s}^2}{2 m_\mu^2}\right]
\left(\frac{2p_\ell}{M_{B_s}}\right)\,.~~~
\eea
Here~\cite{Glashow:2014iga},
\be\label{eq:rhoNP}
\rho_{NP} = \frac{\frac{G}{2}\,U^{d*}_{L33}U^d_{L32}|U^\ell_{L32}|^2}
{ -\frac{4G_F}{\sqrt{2}}V_{tb}^*V_{ts}
\frac{\alpha_{EM}(m_b)}{4\pi}  C_9^e + \frac{G}{2}\,
U^{d*}_{L33}U^d_{L32}|U^\ell_{L32}|^2}
= -0.136\,, 
\ee
where $V = V_{CKM}$, $C_9^e$ is the Wilson coefficient for the operator
$O_{9}$ in $\bar b \to \bar s e^+e^-$, $U^{d *}_{L33} U^d_{L32} \cong
V_{tb}^*V_{ts}$ in the quark-mass model of Ref.~\cite{Martin:2004ec}, and
$p_\ell$ is the momentum of the outgoing lepton in the $B_s$ rest frame.  The
value of $\rho_{NP}$ is obtained from the global-fit result $C_{9,NP} \simeq
- 12\% \, C_{9,SM}$~\cite{Altmannshofer:2014rta}, rather than from $R_{K}$
alone. This $\rho_{NP}$ applies to axial-vector amplitudes as well because
the SM interaction renormalized at $m_b$ is pure LL to a good approximation.

 \begin{table}[!t]
     \begin{center}{
  \begin{tabular}{|c|c|c|c|c|}
  \hline
 Case & $B^+ \to K^+ \mu^\pm \tau^\mp$ & $B^+ \to K^+ e^\pm \tau^\mp$ &
 $B^+ \to K^+ e^\pm \mu^\mp$  &  \\
  \hline\hline
 A & $1.14\times 10^{-8}$ & $3.84\times 10^{-10}$ & $0.52\times 10^{-9}$
 &   \\ 
 D & $0.89\times 10^{-6}$ & $0.67\times 10^{-10}$ & $1.17\times 10^{-12}$
 &   \\
 Exp. & $<4.8\times 10^{-5}$ & $< 3.0\times 10^{-5}$ & $< 9.1\times 10^{-8}$
 & \\
 \hline
 Case & $B_s \to \mu^\pm \tau^\mp$ & $B_s \to e^\pm \tau^\mp$ &
 $B_s \to e^\pm \mu^\mp$  &  $B_s \to \tau^+ \tau^-$ \\
  \hline\hline
 A & $1.37\times 10^{-8}$ & $4.57\times 10^{-10}$ & $1.73 \times 10^{-12}$ & 
     $5.61 \times 10^{-7}$ \\  
 D & $1.06\times 10^{-6}$ & $0.80\times 10^{-10}$ & $3.91\times 10^{-15}$ & 
     $0.76 \times 10^{-4}$ \\
 Exp. & --- & --- & $< 1.1 \times 10^{-8}$ & --- \\
 \hline
 \end{tabular}}
 \caption{Branching ratios for LFV decays of $B$-mesons and $B_s \to
   \tau^+\tau^-$ from Eqs.~(\ref{eq:UlL},\ref{eq:UlR},\ref{eq:LFVBrates},
   \ref{eq:LFVBsrates}), using the central values of $\rho_{NP}$, of
   $\CB(B^+ \to K^+ \mup\mum)\simeq (4.29 \pm 0.22)\times
   10^{-7}$~\cite{Aaij:2014ora} and of $\CB(B_s \to \mup\mum) =
   (2.8^{+0.7}_{-0.6})\times 10^{-9}$~\cite{CMS:2014xfa}. All decays are 
   corrected for phase space (see text). Branching ratio limits are from
   Refs.~\cite{Aaij:2013cby,Agashe:2014kda}.\label{tab:LFVBRs}}
 \label{table}
 \end{center}
 \end{table}

 From Eq.~(\ref{eq:rhoNP}) and the calculated $U_{L}^{d,\ell}$ matrices, one
 can estimate the $G$-coupling strength. For cases~A and~D one has $G \simeq
 4.3 \times 10^{-8}\,\gev^{-2}$ and $1.8 \times 10^{-6}\,\gev^{-2}$. These
 imply the approximate upper bounds $\Lambda_{NP} = 1/\sqrt G = 4.8~\tev$ and
 $745~\gev$, respectively. These mass scales seem low for a $Z'$, but it must
 be remembered that it couples primarily to the third generation.

 There are two approximations in Eq.~(\ref{eq:LFVBrates}) as applied to
 $B \to K \ellp \ellm$ ratios. The denominator is best-measured for $B^+
 \to K^+ \mup\mum$; it is $\CB(B^+ \to K^+ \mu^+\mu^-) = (4.29\pm
 0.22)\times 10^{-7}$ integrated over the full $q^2$-range, $0$--$22
 \,\gev^2$~\cite{Aaij:2014pli}. This integration extrapolates over most of the
 charmonium region, $q^2 = 8$--$15\,\gev^2$, ignoring the presence of the
 narrow resonances. Charmonium resonances do not, of course, influence the
 numerators, but we do not know whether LFV searches will include this
 region. Second, and potentially more important, LFV modes to $\tau$'s have 
 smaller phase space than those with $\mu^+ \mu^-$. For these semileptonic 
 decays, this effect is accounted for using the results of 
 Ref.~\cite{Crivellin:2015era}. 
 Employing our own calculations, we have corrected for phase space the 
 $B_s \to \ell \ell'$ decay rates involving~$\tau$'s. 

 Our results are shown in Table~\ref{table}. As was to be expected for our
 third-generation $\CH_{NP}$, modes involving $\mu\tau$ have the largest
 rates followed by $e\tau$ with rates smaller by one or more orders of
 magnitude. Rates for the experimentally easier $e\mu$ modes are very small
 and may be beyond reach in the near future; the one exception in our model
 is $B^+ \to K^+ e\mu$ in case~A. The large $B_s \to \mu\tau$ rate
 predictions are not yet excluded. The best public limits on these LFV modes
 are also listed in Table~\ref{table}~\cite{Aaij:2013cby,Agashe:2014kda}.

In conclusion, it is natural to ask how general are our results; are they to
be expected in other NP models of the $B$-decay anomalies or in other schemes
for calculating the mixing matrices? Particularly, is the relative importance
of $B \to X\mu\tau$ that we found likely to be a common feature of such
models? It is hard to be sure, of course, but we do believe it is. As we
emphasized of our $\CH_{NP}$, the LHCb data strongly points to the third
generation, or at most just the second and third generations, as the seat of
lepton nonuniversality.  Further, the hierarchy of charged-lepton masses ---
not unlike that for the quarks --- suggests block-diagonal mass matrices and,
therefore, mixing matrices somewhere along the line from our original
CKM-like expectation to the ones we found in Eqs.~(\ref{eq:UlL},\ref{eq:UlR})
from the ABP ansatz.

These expectations can be compared with those obtained within other proposed 
flavor models. Our prediction of a generic enhancement over the SM rate of decay 
modes involving the third generation is also advertised in the class of models 
discussed in Ref.~\cite{Alonso:2015sja}, although they have unobservable LFV by 
construction. The only model allowing for a direct comparison is 
Ref.~\cite{Boucenna:2015raa}. For the $B \to K$ transitions to either $e \tau$ 
or $e \mu$, the branching-ratio ranges predicted in our cases A and D encompass 
those predicted in their model. In the $\mu \tau$ case our predictions are above 
theirs in both A and D cases, although they also predict a relative enhancement 
of this channel with respect to the other LFV modes.

Therefore, in addition to more firmly establishing the
apparent lepton nonuniversality in $B$ decays, it is important that LHCb and
other experiments mount searches for lepton flavor violation, with special
attention to improving significantly the limits on $\mu\tau$ and even $e\tau$
decay modes.

\section*{Acknowledgments}

We thank Tom Appelquist, Aoife Bharucha, Cedric Delaunay, Shelly Glashow, 
David London, Marco Nardecchia, Maurizio Piai, David Straub and Edwige Tournefier 
for stimulating and helpful conversations.  KL gratefully acknowledges support of 
this project by a CERN
Scientific Associateship and by the Labex ENIGMASS. He thanks the CERN Theory
Group and Laboratoire d'Annecy-le-Vieux de Physique Th\'eorique (LAPTh) for
their gracious hospitality in 2014-15. KL's research is also supported by the
U.S.~Department of Energy under Grant No.~DE-SC0010106.


\bibliography{LFV2}

\providecommand{\href}[2]{#2}\begingroup\raggedright\begin{thebibliography}{10}

\bibitem{Aaij:2014ora}
{\bf LHCb} Collaboration, R.~Aaij {\em et.~al.}, ``{Test of lepton universality
  using $B^{+}\rightarrow K^{+}\ell^{+}\ell^{-}$ decays},'' {\em
  Phys.Rev.Lett.} {\bf 113} (2014) 151601,
  \href{http://xxx.lanl.gov/abs/1406.6482}{ 1406.6482}.

\bibitem{Bobeth:2007dw}
C.~Bobeth, G.~Hiller, and G.~Piranishvili, ``{Angular distributions of $\bar B
  \to K \bar \ell \ell$ decays},'' {\em JHEP} {\bf 0712} (2007) 040,
  \href{http://xxx.lanl.gov/abs/0709.4174}{ 0709.4174}.

\bibitem{Bouchard:2013mia}
{\bf HPQCD} Collaboration, C.~Bouchard, G.~P. Lepage, C.~Monahan, H.~Na, and
  J.~Shigemitsu, ``{Standard Model Predictions for $B \to K \ell^+ \ell^-$ with
  Form Factors from Lattice QCD},'' {\em Phys.Rev.Lett.} {\bf 111} (2013),
  no.~16, 162002, \href{http://xxx.lanl.gov/abs/1306.0434}{ 1306.0434}.

\bibitem{Hiller:2003js}
G.~Hiller and F.~Kruger, ``{More model independent analysis of $b \to s$
  processes},'' {\em Phys.Rev.} {\bf D69} (2004) 074020,
  \href{http://xxx.lanl.gov/abs/hep-ph/0310219}{ hep-ph/0310219}.

\bibitem{Aaij:2014pli}
{\bf LHCb} Collaboration, R.~Aaij {\em et.~al.}, ``{Differential branching
  fractions and isospin asymmetries of $B \to K^{(*)} \mu^+ \mu^-$ decays},''
  {\em JHEP} {\bf 1406} (2014) 133, \href{http://xxx.lanl.gov/abs/1403.8044}{
  1403.8044}.

\bibitem{Bobeth:2011gi}
C.~Bobeth, G.~Hiller, and D.~van Dyk, ``{More Benefits of Semileptonic Rare B
  Decays at Low Recoil: CP Violation},'' {\em JHEP} {\bf 1107} (2011) 067,
  \href{http://xxx.lanl.gov/abs/1105.0376}{ 1105.0376}.

\bibitem{Bobeth:2011nj}
C.~Bobeth, G.~Hiller, D.~van Dyk, and C.~Wacker, ``{The Decay $B \to K \ell^+
  \ell^-$ at Low Hadronic Recoil and Model-Independent $\Delta B = 1$
  Constraints},'' {\em JHEP} {\bf 1201} (2012) 107,
  \href{http://xxx.lanl.gov/abs/1111.2558}{ 1111.2558}.

\bibitem{Bobeth:2012vn}
C.~Bobeth, G.~Hiller, and D.~van Dyk, ``{General analysis of $\bar{B} \to
  \bar{K}^{(*)}\ell^+ \ell^-$ decays at low recoil},'' {\em Phys.Rev.} {\bf
  D87} (2013), no.~3, 034016, \href{http://xxx.lanl.gov/abs/1212.2321}{
  1212.2321}.

\bibitem{LHCb:2015iha}
{\bf LHCb} Collaboration, C.~Langenbruch, ``{Latest results on rare decays from
  LHCb},'' \href{http://xxx.lanl.gov/abs/1505.04160}{ 1505.04160}.

\bibitem{Descotes-Genon:2013wba}
S.~Descotes-Genon, J.~Matias, and J.~Virto, ``{Understanding the $B \to K^{*}
  \mu^{+} \mu^{-}$ Anomaly},'' {\em Phys.Rev.} {\bf D88} (2013) 074002,
  \href{http://xxx.lanl.gov/abs/1307.5683}{ 1307.5683}.

\bibitem{Descotes-Genon:2014uoa}
S.~Descotes-Genon, L.~Hofer, J.~Matias, and J.~Virto, ``{On the impact of power
  corrections in the prediction of $B \to K^*\mu^+\mu^-$ observables},'' {\em
  JHEP} {\bf 1412} (2014) 125, \href{http://xxx.lanl.gov/abs/1407.8526}{
  1407.8526}.

\bibitem{Jager:2014rwa}
S.~Jaeger and J.~Martin~Camalich, ``{Reassessing the discovery potential of the
  $B \to K^{*} \ell^+\ell^-$ decays in the large-recoil region: SM challenges
  and BSM opportunities},'' \href{http://xxx.lanl.gov/abs/1412.3183}{
  1412.3183}.

\bibitem{CMS:2014xfa}
{\bf CMS, LHCb} Collaboration, V.~Khachatryan {\em et.~al.}, ``{Observation of
  the rare $B^0_s\to\mu^+\mu^-$ decay from the combined analysis of CMS and
  LHCb data},'' {\em Nature} (2015) \href{http://xxx.lanl.gov/abs/1411.4413}{
  1411.4413}.

\bibitem{Bobeth:2013uxa}
C.~Bobeth, M.~Gorbahn, T.~Hermann, M.~Misiak, E.~Stamou, {\em et.~al.},
  ``{$B_{s,d} \to \ell^+ \ell^-$ in the Standard Model with Reduced Theoretical
  Uncertainty},'' {\em Phys.Rev.Lett.} {\bf 112} (2014) 101801,
  \href{http://xxx.lanl.gov/abs/1311.0903}{ 1311.0903}.

\bibitem{Glashow:2014iga}
S.~L. Glashow, D.~Guadagnoli, and K.~Lane, ``{Lepton Flavor Violation in $B$
  Decays?},'' {\em Phys.Rev.Lett.} {\bf 114} (2015) 091801,
  \href{http://xxx.lanl.gov/abs/1411.0565}{ 1411.0565}.

\bibitem{Altmannshofer:2013foa}
W.~Altmannshofer and D.~M. Straub, ``{New physics in $B \to K^*\mu\mu$?},''
  {\em Eur.Phys.J.} {\bf C73} (2013) 2646,
  \href{http://xxx.lanl.gov/abs/1308.1501}{ 1308.1501}.

\bibitem{Gauld:2013qba}
R.~Gauld, F.~Goertz, and U.~Haisch, ``{On minimal $Z'$ explanations of the $B
  \to K^{*} \mu^{+} \mu^{-}$ anomaly},'' {\em Phys.Rev.} {\bf D89} (2014)
  015005, \href{http://xxx.lanl.gov/abs/1308.1959}{ 1308.1959}.

\bibitem{Buras:2013qja}
A.~J. Buras and J.~Girrbach, ``{Left-handed Z' and Z FCNC quark couplings
  facing new $b \to s \mu^+ \mu^-$ data},'' {\em JHEP} {\bf 1312} (2013) 009,
  \href{http://xxx.lanl.gov/abs/1309.2466}{ 1309.2466}.

\bibitem{Gauld:2013qja}
R.~Gauld, F.~Goertz, and U.~Haisch, ``{An explicit Z'-boson explanation of the
  $B \to K^* \mu^+ \mu^-$ anomaly},'' {\em JHEP} {\bf 1401} (2014) 069,
  \href{http://xxx.lanl.gov/abs/1310.1082}{ 1310.1082}.

\bibitem{Buras:2013dea}
A.~J. Buras, F.~De~Fazio, and J.~Girrbach, ``{331 models facing new $b \to
  s\mu^+ \mu^-$ data},'' {\em JHEP} {\bf 1402} (2014) 112,
  \href{http://xxx.lanl.gov/abs/1311.6729}{ 1311.6729}.

\bibitem{Altmannshofer:2014cfa}
W.~Altmannshofer, S.~Gori, M.~Pospelov, and I.~Yavin, ``{Quark flavor
  transitions in $L_\mu-L_\tau$ models},'' {\em Phys.Rev.} {\bf D89} (2014)
  095033, \href{http://xxx.lanl.gov/abs/1403.1269}{ 1403.1269}.

\bibitem{Hiller:2014yaa}
G.~Hiller and M.~Schmaltz, ``{$R_K$ and future $b \to s \ell \ell$ physics
  beyond the standard model opportunities},'' {\em Phys.Rev.} {\bf D90} (2014)
  054014, \href{http://xxx.lanl.gov/abs/1408.1627}{ 1408.1627}.

\bibitem{Ghosh:2014awa}
D.~Ghosh, M.~Nardecchia, and S.~Renner, ``{Hint of Lepton Flavour
  Non-Universality in $B$ Meson Decays},'' {\em JHEP} {\bf 1412} (2014) 131,
  \href{http://xxx.lanl.gov/abs/1408.4097}{ 1408.4097}.

\bibitem{Altmannshofer:2014rta}
W.~Altmannshofer and D.~M. Straub, ``{New physics in $b \to s$ transitions
  after LHC run 1},'' \href{http://xxx.lanl.gov/abs/1411.3161}{ 1411.3161}.

\bibitem{Hiller:2014ula}
G.~Hiller and M.~Schmaltz, ``{Diagnosing lepton-nonuniversality in $b \to s
  \ell \ell$},'' {\em JHEP} {\bf 1502} (2015) 055,
  \href{http://xxx.lanl.gov/abs/1411.4773}{ 1411.4773}.

\bibitem{Gripaios:2014tna}
B.~Gripaios, M.~Nardecchia, and S.~Renner, ``{Composite leptoquarks and
  anomalies in $B$-meson decays},'' {\em JHEP} {\bf 1505} (2015) 006,
  \href{http://xxx.lanl.gov/abs/1412.1791}{ 1412.1791}.

\bibitem{Bhattacharya:2014wla}
B.~Bhattacharya, A.~Datta, D.~London, and S.~Shivashankara, ``{Simultaneous
  Explanation of the $R_K$ and $R(D^{(*)})$ Puzzles},'' {\em Phys.Lett.} {\bf
  B742} (2015) 370--374, \href{http://xxx.lanl.gov/abs/1412.7164}{ 1412.7164}.

\bibitem{Crivellin:2015mga}
A.~Crivellin, G.~D'Ambrosio, and J.~Heeck, ``{Explaining $h\to\mu^\pm\tau^\mp$,
  $B\to K^* \mu^+\mu^-$ and $B\to K \mu^+\mu^-/B\to K e^+e^-$ in a
  two-Higgs-doublet model with gauged $L_\mu-L_\tau$},'' {\em Phys.Rev.Lett.}
  {\bf 114} (2015) 151801, \href{http://xxx.lanl.gov/abs/1501.00993}{
  1501.00993}.

\bibitem{Crivellin:2015lwa}
A.~Crivellin, G.~D'Ambrosio, and J.~Heeck, ``{Addressing the LHC flavor
  anomalies with horizontal gauge symmetries},'' {\em Phys.Rev.} {\bf D91}
  (2015), no.~7, 075006, \href{http://xxx.lanl.gov/abs/1503.03477}{
  1503.03477}.

\bibitem{Niehoff:2015bfa}
C.~Niehoff, P.~Stangl, and D.~M. Straub, ``{Violation of lepton flavour
  universality in composite Higgs models},'' {\em Phys.Lett.} {\bf B747} (2015)
  182--186, \href{http://xxx.lanl.gov/abs/1503.03865}{ 1503.03865}.

\bibitem{Sierra:2015fma}
S.~D. Aristizabal, F.~Staub, and A.~Vicente, ``{Shedding light on the $b\to s$
  anomalies with a dark sector},'' \href{http://xxx.lanl.gov/abs/1503.06077}{
  1503.06077}.

\bibitem{Celis:2015ara}
A.~Celis, J.~Fuentes-Martin, M.~Jung, and H.~Serodio, ``{Family non-universal
  Z' models with protected flavor-changing interactions},''
  \href{http://xxx.lanl.gov/abs/1505.03079}{ 1505.03079}.

\bibitem{Becirevic:2015asa}
D.~Becirevic, S.~Fajfer, and N.~Kosnik, ``{Lepton flavor non-universality in $b
  \to s \ell^+ \ell^-$ processes},'' \href{http://xxx.lanl.gov/abs/1503.09024}{
  1503.09024}.

\bibitem{Varzielas:2015iva}
I.~de~Medeiros~Varzielas and G.~Hiller, ``{Clues for flavor from rare lepton
  and quark decays},'' {\em JHEP} {\bf 1506} (2015) 072,
  \href{http://xxx.lanl.gov/abs/1503.01084}{ 1503.01084}.

\bibitem{Boucenna:2015raa}
S.~M. Boucenna, J.~W.~F. Valle, and A.~Vicente, ``{Are the B decay anomalies
  related to neutrino oscillations?},''
  \href{http://xxx.lanl.gov/abs/1503.07099}{ 1503.07099}.

\bibitem{Crivellin:2015era}
A.~Crivellin, L.~Hofer, J.~Matias, U.~Nierste, S.~Pokorski, {\em et.~al.},
  ``{Lepton-Flavour Violating $B$ Decays in Generic $Z^\prime$ Models},''
  \href{http://xxx.lanl.gov/abs/1504.07928}{ 1504.07928}.

\bibitem{Lee:2015qra}
C.-J. Lee and J.~Tandean, ``{Minimal Lepton Flavor Violation Implications of
  the $b \to s$ Anomalies},'' \href{http://xxx.lanl.gov/abs/1505.04692}{
  1505.04692}.

\bibitem{Alonso:2015sja}
R.~Alonso, B.~Grinstein, and J.~M. Camalich, ``{Lepton universality violation
  and lepton flavor conservation in $B$-meson decays},''
  \href{http://xxx.lanl.gov/abs/1505.05164}{ 1505.05164}.

\bibitem{Greljo:2015mma}
A.~Greljo, G.~Isidori, and D.~Marzocca, ``{On the breaking of Lepton Flavor
  Universality in B decays},'' \href{http://xxx.lanl.gov/abs/1506.01705}{
  1506.01705}.

\bibitem{Calibbi:2015kma}
L.~Calibbi, A.~Crivellin, and T.~Ota, ``{Effective field theory approach to
  $b\to s\ell\ell^{(\prime)}$, $B\to K^{(*)}\nu\bar{\nu}$ and $B\to
  D^{(*)}\tau\nu$ with third generation couplings},''
  \href{http://xxx.lanl.gov/abs/1506.02661}{ 1506.02661}.

\bibitem{Hurth:2014vma}
T.~Hurth, F.~Mahmoudi, and S.~Neshatpour, ``{Global fits to $b \to s\ell\ell$
  data and signs for lepton non-universality},'' {\em JHEP} {\bf 1412} (2014)
  053, \href{http://xxx.lanl.gov/abs/1410.4545}{ 1410.4545}.

\bibitem{Alonso:2014csa}
R.~Alonso, B.~Grinstein, and J.~Martin~Camalich, ``{$SU(2)\times U(1)$ gauge
  invariance and the shape of new physics in rare $B$ decays},'' {\em
  Phys.Rev.Lett.} {\bf 113} (2014) 241802,
  \href{http://xxx.lanl.gov/abs/1407.7044}{ 1407.7044}.

\bibitem{Agashe:2014kda}
{\bf Particle Data Group} Collaboration, K.~Olive {\em et.~al.}, ``{Review of
  Particle Physics},'' {\em Chin.Phys.} {\bf C38} (2014) 090001.

\bibitem{Appelquist:2015mga}
T.~Appelquist, Y.~Bai, and M.~Piai, ``{Constrained Flavor Breaking},'' {\em
  Phys.Rev.} {\bf D91} (2015), no.~9, 093009,
  \href{http://xxx.lanl.gov/abs/1503.07450}{ 1503.07450}.

\bibitem{Feruglio:2015jfa}
F.~Feruglio, ``{Pieces of the Flavour Puzzle},''
  \href{http://xxx.lanl.gov/abs/1503.04071}{ 1503.04071}.

\bibitem{Martin:2004ec}
A.~Martin and K.~Lane, ``{CP violation and flavor mixing in technicolor
  theories},'' {\em Phys.Rev.} {\bf D71} (2005) 015011,
  \href{http://xxx.lanl.gov/abs/hep-ph/0404107}{ hep-ph/0404107}.

\bibitem{Harari:1986xf}
H.~Harari and M.~Leurer, ``{Recommending a Standard Choice of Cabibbo Angles
  and KM Phases for Any Number of Generations},'' {\em Phys.Lett.} {\bf B181}
  (1986) 123.

\bibitem{CKM}
CKMfitter Group (J. Charles {\em et al.}), Eur. Phys. J. C41, 1-131 (2005)
  [hep-ph/0406184], updated results and plots available at:
  http://ckmfitter.in2p3.fr. UTfit Group (M. Ciuchini {\em et al.}), JHEP {\bf
  0107} (2001) 013 [hep-ph/0012308], updated results and plots available at:
  http://www.utfit.org.

\bibitem{Aaij:2013cby}
{\bf LHCb} Collaboration, R.~Aaij {\em et.~al.}, ``{Search for the
  lepton-flavor violating decays $B^0_s \rightarrow e^{\pm}\mu^{\mp}$ and $B^0
  \rightarrow e^{\pm} \mu^{\mp}$},'' {\em Phys.Rev.Lett.} {\bf 111} (2013)
  141801, \href{http://xxx.lanl.gov/abs/1307.4889}{ 1307.4889}.

\end{thebibliography}\endgroup
\bibliographystyle{utcaps}
\end{document}